# Deployment of software components: Application to Wireless System

KOUNINEF Belkacem[1], BOUZERITA Mohamed[2]

Institut National des Télécommunications et des
Technologies de l'Information et de la Communication
Oran, Algérie
[1]bkouninef@ito.dz, [2]bouzerita_tic@yahoo.fr

**Abstract.** The wide variety of wireless devices brings to design mobile applications as a collection of interchangeable software components adapted to the deployment environment of the software. To ensure the proper functioning of the software assembly and make a real enforcement in case of failures, the introduction of concepts, models and tools necessary for the administration of these components is crucial. This article proposes a method for deploying components in wireless systems.

**Mots-clés :** J2ME, SyncML, OMA DM.

## 1. Introduction

Wireless systems, such as PDAs, wireless phones and on-board computers in cars can now have permanent access to a large number of applications: mail, word processing, GPS navigation system, m-learning, photo applications, video applications, etc.. The deployments of these applications are becoming increasingly complex due to the diversity of access terminals and communication infrastructure. It is based on highly centralized tools and many manual operations for the administrator. These techniques are not really usable in the network without administrator (network ad-hoc).

Platforms deployment starts to make deployment easier on desktop machines. The first standardization work (under the OMA[2]) have permitted the emergence of specifications for deployment. In this context, we would build a prototype platform deployment that would deploy a local (peer to peer) application on a wireless system (mobile phones).

## 2. Software component and Wireless system

Wireless systems have become more complex and numerous approaches take advantage of today's software components paradigm to develop applications embedded in these systems[6].

J2ME [5] defines a model of components (MIDlets). The life cycle of these components is managed by software residing on the wireless system, AMS (Application Management System). This management is crucial for wireless systems because they have limited resources.

## 3. OMA Device Management Protocol

OMA is a large organization of mobile phone manufacturers, which is responsible for managing standards for portable equipment. Based on the SyncML protocol [1] in 2004, the AMO has proposed two standard services: data synchronization (Data Synchronization - SD) and management of devices (Device Management - DM). The OMA DM [3] is a protocol that allows operating and configuring the equipment, access and control resources on the mobile. An advantage of the OMA DM is the ability to deploy applications. In this section, we present the protocol for management of OMA applications. It consists of two parts:

- The model data available for remote handling.
- The communication protocol between the management server and mobile client.

### 3.1 Data Model

It is difficult for the server to manage mobiles because of their configuration and their specific characteristics that depend on each manufacturer [7]. OMA DM has proposed a standard framework to enable providers to identify the description of devices and communicate it to the server. The server is based on the description to send operations. This description is called "the tree of device management".

*Tree management devices*

Figure 1 depicts an example of a tree management device. This tree holds all the managed objects of the device as a tree structure where each node is addressed (uniquely) by a URI (*Uniform Resource Identifier*).

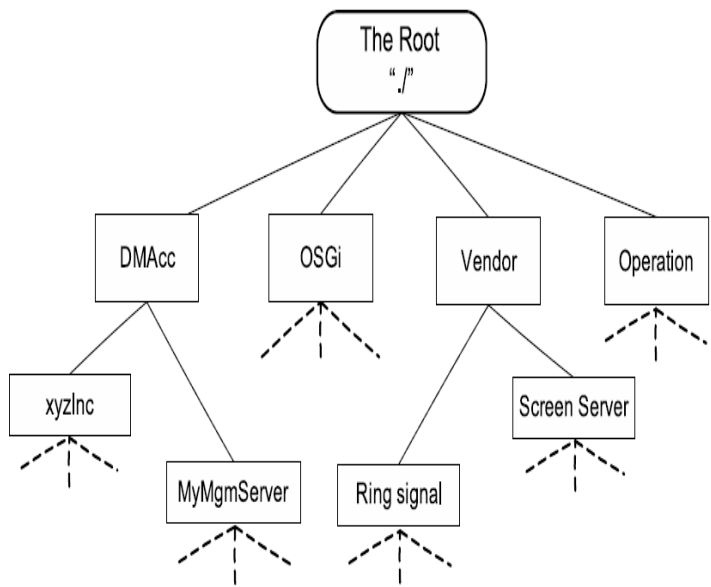

Figure1. Tree management device.

The node can take any form: a single parameter, a GIF or even a complete application. There are two types of objects:
- The permanent objects are incorporated into the device, they cannot be deleted. For example, the DevInfo object that specifies the basic information of the device.
- Dynamic objects are objects that can be added or deleted. (For example, tones).

*Features Node*

Nodes are entities that can be manipulated by the OMA DM protocol. Each tree node has a set of properties defining characteristics:

- ACL (Access Control List) defines who can use the node.
- Format: Determines how a node is interpreted.
- Name: the name of the node.
- Size: The size of the information the node contains.
- Title: the name of the node visible.
- Timestamp: date and time of its last modification.
- Type: the MIME type of information the node contains.
- Verno: the version number, automatically incremented with each modification.

*Nodes Manipulation*

The management nodes can be handled by SyncML DM messages containing the following commands:

- Get: allows the server management to explore the structure of the tree. If the node is accessed within a node, a list of all node names is returned and if the node is a leaf, its information is returned.
- Add: Adds a node to a tree.
- Replace: Replaces a node in the tree.
- Delete: Deletes a node from the tree.
- Copy: Copies a node of the tree.

The server can modify the tree management during the execution using Add commands. The new node can be an internal or a leaf node but the parent must be an interior node. The device itself can also change the tree.

*Standard Objects*

OMA DM proposes a framework that allows manufacturers to set their own descriptions for their products. But it would be limited if the entities in the manageable devices have different formats and different behaviors. To avoid this, the OMA has defined a number of items required. These objects are mainly associated with OMA DM [3] and the configuration of SyncML:
- DevInfo description of the device, sent from client to server.
- DevDetail: General information about the device that benefits from standardization.
- DMAcc: Institution for the DM client.

**3.2 Protocol and mechanism**
OMA provides a communication protocol between the client and the server for device management. The protocol consists of a sequence of messages between the client and the server to perform operations.

*Protocol packets*

The protocol consists of two main phases, the initiation phase and the management phase. The first phase is designed to authenticate and exchange information of the device. The second makes commands of the server on the client. These phases are considered as transactions of the SyncML packets.

The management phase consists of a number of iterations. The contents of the packet sent by the server determine whether the session should be continued or not. If the server sends the transactions in a package that needs a response (Status or Results), this phase continues with a new client package for the server containing the answers. The response packet of the client starts a new iteration. The server may send a new package management operation and then starts a new iteration if desired [13]. The processing of packets can make an unpredictable time. Therefore, the OMA DM protocol indicates no timeout between packets.

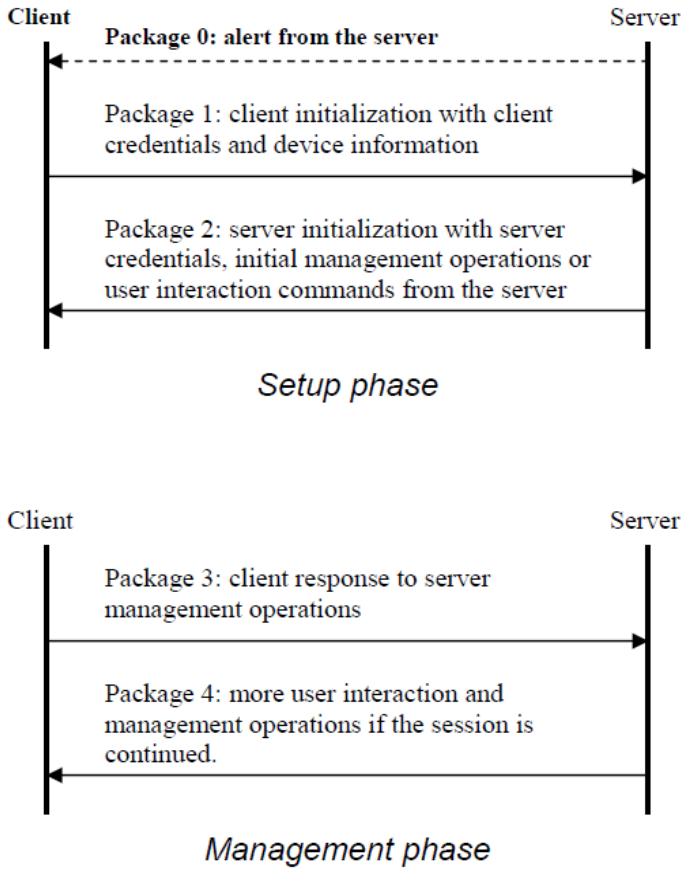

Figure 2. Packages of the OMA DM protocol.

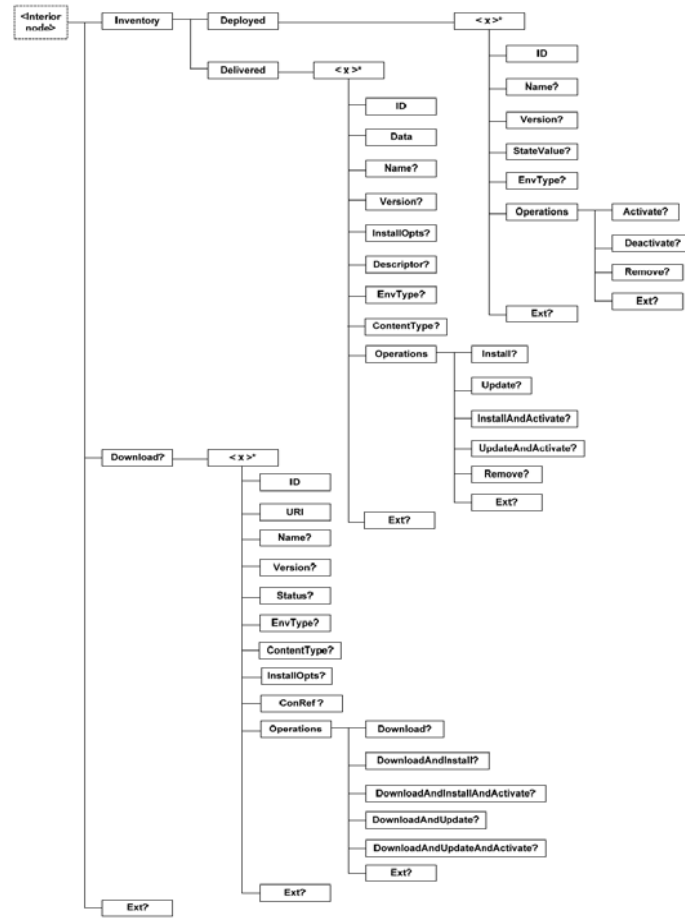

Figure 3. Application management Object description [4].

## 4. Deployment system

This system creates an arborescence on the tree management device. This structure is the basis on which management can make all applications. It makes the management of the application easier.

### 4.1 Application management Object description

The application management object provides functionalities for remote application management; node in the internal management object is called SCM (Software Component Management) [11] and can be found then from the root node (Figure 3).

- Delivery of Application: send an application from the server to the device. The consignment consists of information on the application (such as name, version, etc...) Description file. Jad and the execution file. Applications delivered by the OMA DM server are located in the interior node Inventory / Delivered. Be aware of the type and the size of the mobile application, especially the MIDlet.

- Inventory of a device: the discovery of device applications. In fact, all the nodes of the tree management are taken and classified. This process sends to the server a list of applications and their status. The application state depends on its position on the tree. The server is based on that state to decide what action can be executed with the appropriate application. To perform any action deployment, we must first identify the device.

- Update of an application: Exchange of information on an application and the runtime files. We can only update applications that have been delivered by the OMA DM server [3].

- Installing an application: when you deliver applications to the device, they cannot be activated. To activate, you must first install. Indeed, each node contains a set of commands for application deployment. In this case, to install an application, we send the command Exec to operation node appropriate to the object that represents the application.

- Application download: send application information with its URI (file. Jad or. Jar). Objects that describe downloadable applications are in the interior node Dowload. To start the download, the server sends the command Exec to node operation appropriate to the subject.

- Enabling / disabling an application: the command Exec is sent to the operation node appropriate for the object that represents the application.

- Deleting an Application: Removes all nodes on the application on the management tree.

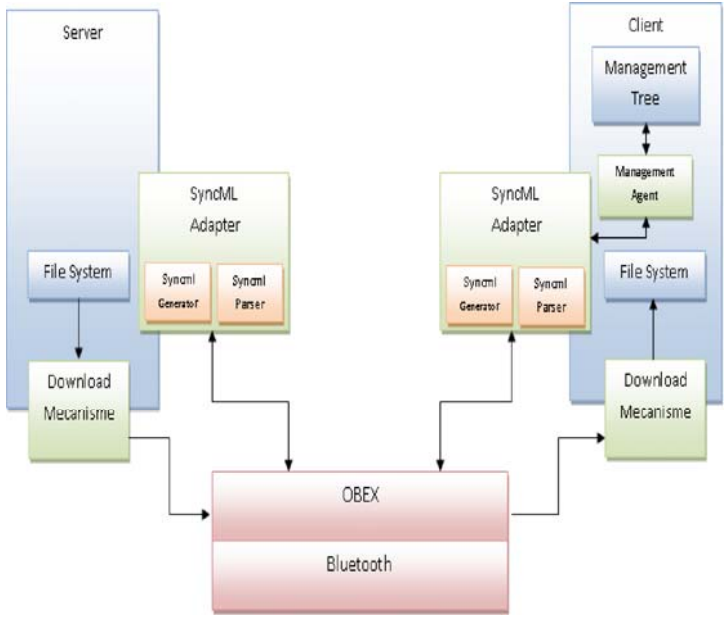

Figure4. Architecture of the Deployment System (prototype) peer - peer according to the OMA DM (OMA DM over OBEX / Bluetooth).

- SyncML Adapter
  a. SyncML Genenrator(Serevr) : generates management commands as SyncML messages.
  b. SyncML Parser(Client) : SyncML analyzes massages to extract management commands.
- Management Tree: configuration file in XML format (Figure 2).
- Management Agent: XML parser (DOM) enables the creation; modification and removal of elements of the configuration file (Tree Management).
- Download Macanisme : FTP OBEX.
- OBEX/Bluetooth : Transport Layer.

OMA DM Client:
1: Server Bluetooth Address
2: Name of service
3: The root element (Software Component Management)
4: Structure of the Tree Management Application
5: Messages syncml Server to Client
6: Data from one node (descriptor)

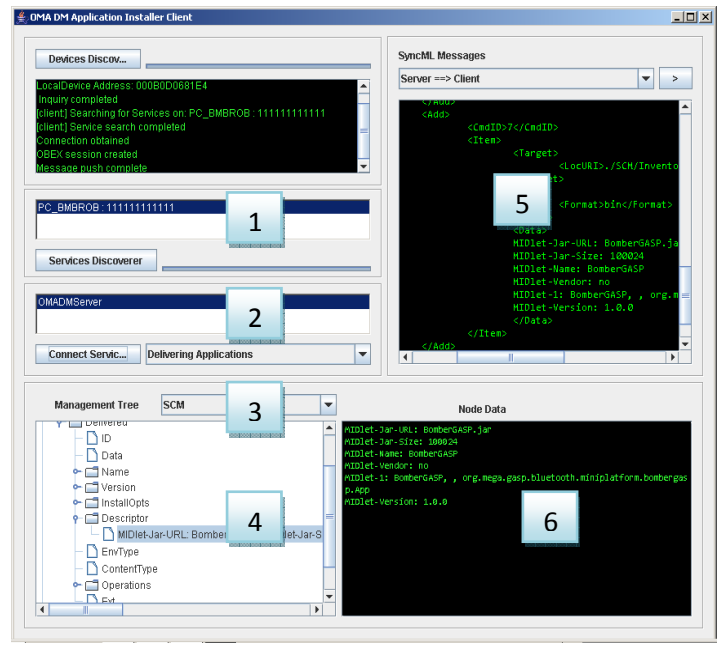

Figure5. Client interface.

OMA DM Server:
1: SyncML Parser
2: Message SyncML Server to Client
3: Name and address of the client's bluetooth
4: Structure of the Tree Management (DivInfo)

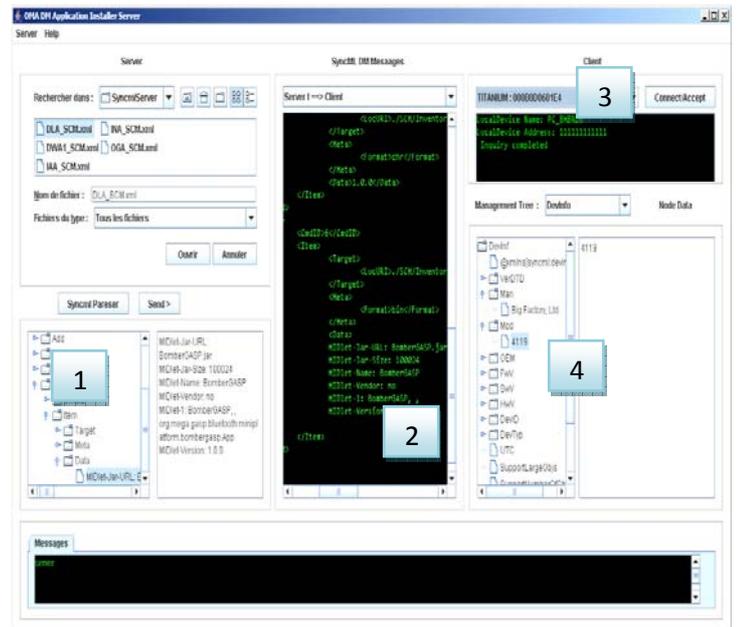

Figure 6. Server interface.

## 5. CONCLUSION

The wide variety of wireless devices brings to design mobile applications as an assembly of interchangeable software components adapted to the deployment environment of the software [8]. To ensure the proper functioning of the software assembly and make a real enforcement in case of failures, the establishment of concepts, models and tools necessary to manage these components is necessary[13].
J2ME defines a model of components (MIDlets). The life cycle of these components is managed by software residing on the wireless system, AMS (Application Management System). This management is crucial for wireless systems because they have limited resources.

OMA is a large organization of mobile phone manufacturers, which is responsible for managing standards for portable equipment. Based on the SyncML protocol in 2004, the AMO has proposed two standard services: data synchronization (DS - Data Synchronization) and management of devices (Device Management-DM) [12].

The OMA DM is a protocol that allows exploiting and configuring devices, access and controlling resources on mobiles. One advantage of the OMA DM is the ability to deploy applications.

In this paper, architecture for deploying software components in wireless systems was presented. The objective was to develop a system for local deployment for mobile devices; this system must take into account all deployment activities: [10] installation, upgrade, activation, deactivation and uninstalling.

To focus on our problem of deploying components, we have built our approach on a simple component model based on MIDlet, it would be interesting to apply our solution to a component model like OSGi [9], which is becoming more and more popular in the Wireless World.